\documentclass[10pt]{article}
\usepackage{times,url}
\usepackage{array,psfig,epsf,epsfig}
\newtheorem{theorem}{Theorem}[section]

\begin{document}
\bibliographystyle{plain}

\title{sSCADA: Securing SCADA Infrastructure Communications}

\author{Yongge Wang \\ Dept. of SIS, UNC Charlotte,\\
9201 University City Blvd, Charlotte, NC 28223
{\tt yonwang@uncc.edu}}


\setcounter{page}{1}
\maketitle

\begin{abstract}
Distributed control systems (DCS) and supervisory control and data 
acquisition (SCADA) systems were developed to reduce labor costs, and 
to allow system-wide monitoring and remote control from a central location. 
Control systems are widely used in critical infrastructures such as electric 
grid, natural gas, water, and wastewater industries. 
While control systems can be vulnerable to a variety of types of cyber 
attacks that could have devastating consequences, little research
has been done to secure the control systems. American Gas Association (AGA),
IEC TC57 WG15, IEEE, NIST, and National SCADA Test Bed Program have been 
actively designing cryptographic standard to protect SCADA systems.
American Gas Association (AGA) had originally been designing 
cryptographic standard to protect SCADA communication links and finished
the report AGA 12 part 1. The AGA 12 part 2 has been trransferred to 
IEEE P1711. This paper presents an attack on the protocols in the first draft
of AGA standard \cite{acns}. This attack shows that the security mechanisms
in the first version of the AGA standard protocol could be easily defeated. 
We then propose a suite
of security protocols optimized for SCADA/DCS systems which include:
point-to-point secure channels, authenticated broadcast channels, 
authenticated emergency channels, and revised authenticated 
emergency channels. These protocols are designed to address the 
specific challenges that SCADA systems have.
\end{abstract}

\section{Introduction}
Control systems are computer-based systems that are used within
many critical infrastructures and industries (e.g., electric grid, 
natural gas, water, and wastewater industries)
to monitor and control sensitive processes and physical functions. 
Without a secure SCADA system it is impossible to protect the nation's 
critical infrastructures. 

Typically, control systems collect sensor measurements and operational 
data from the field, process and display this information, and relay 
control commands to local or remote equipments. Control systems may perform 
additional control functions such as operating railway switches, 
circuit breakers, and adjusting valves to regulate flow in pipelines. 
The most sophisticated ones control devices and systems at an even higher 
level. 

Control systems have been in place since the 1930s and 
there are two primary types of control systems. 
Distributed Control Systems (DCS) and Supervisory Control and Data 
Acquisition (SCADA) systems. DCS systems typically are used within a single
processing or generating plant or over a small geographic area. SCADA 
systems typically are used for large, geographically
dispersed distribution operations. For example, a utility company may
use a DCS to generate power and a SCADA system to distribute it.
We will concentrate on SCADA systems and our discussions are
generally applicable to DCS systems. 

\begin{center}
\begin{figure}[htb]
\centering{\includegraphics[scale=0.8]{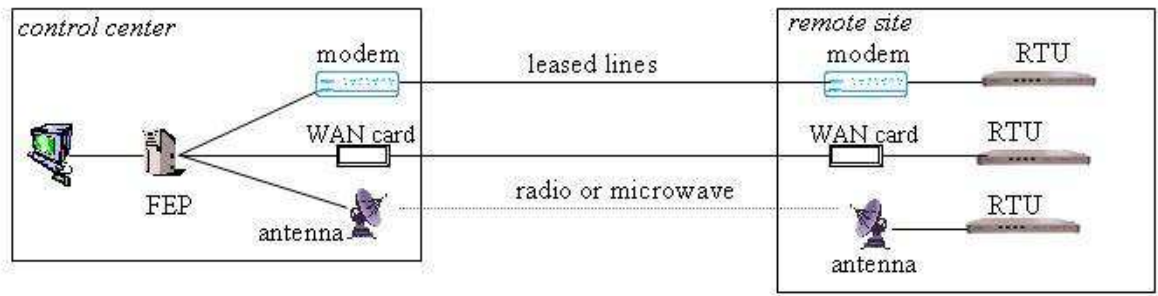}}
\caption{A simple SCADA system}
\label{scadapic}
\end{figure}
\end{center}

\begin{center}
\begin{figure}[htb]
\centering{\includegraphics[scale=0.8]{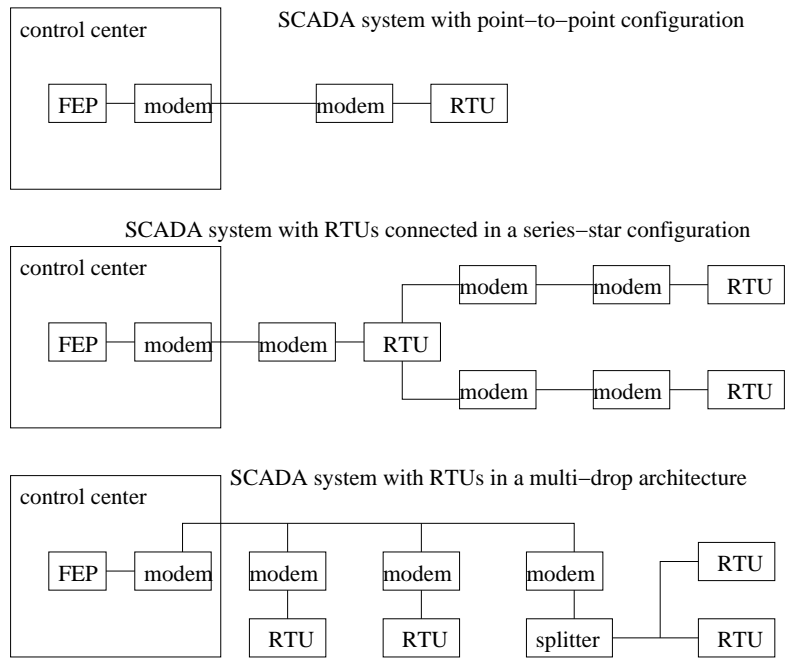}}
\caption{Typical SCADA system configurations}
\label{scadaconfig}
\end{figure}
\end{center}

In a typical SCADA system \cite{scadabook}, data acquisition and control 
are performed by remote terminal units (RTU) and field devices 
that include functions for communications and signaling. SCADA systems 
normally use a poll-response model for communications with clear text 
messages. Poll messages are typically small (less than 16 bytes) and responses
might range from a short ``I am here'' to a dump of an entire day's data.
Some SCADA systems may also allow for unsolicited reporting from remote units.
The communications between the control center and remote sites
could be classified into following four categories.
\begin{enumerate}
\item {\em Data acquisition}: the control center sends poll (request) 
messages to remote terminal units (RTU) and the RTUs dump data to 
the control center. In particular, this includes
{\em status scan and measured value scan}. The control center
regularly sends a status scan request to remote sites 
to get field devices status (e.g., OPEN or CLOSED or 
a fast CLOSED-OPEN-CLOSED sequence) and a measured value scan request
to get measured values of field devices.
The measured values could be analog values or digitally coded values
and are scaled into engineering format by the front-end processor (FEP) 
at the control center.
\item {\em Firmware download}: the control center  sends firmware downloads 
to remote sites. In this case, the poll message is larger 
(e.g., larger than 64K bytes) than other cases.
\item {\em Control functions}: the control center sends control commands
to a RTU at remote sites. Control functions are grouped into 
four subclasses:
individual device control (e.g., to turn on/off a remote device),
control messages to regulating equipment (e.g., a RAISE/LOWER command 
to adjust the remote valves), 
sequential control schemes (a series of correlated individual control 
commands), 
and automatic control schemes (e.g., closed control loops).
\item {\em Broadcast}: the control center may broadcast messages
to multiple remote terminal units (RTUs). For example, the control center
broadcasts an emergent shutdown message or a set-the-clock-time message.
\end{enumerate}
Acquired data is automatically monitored at the control center
to ensure that measured and calculated values lie within permissible limits.
The measured values are monitored with regard to rate-of-change and
for continuous trend monitoring. They are also recorded for post-fault
analysis. Status indications are monitored at the control center
with regard to changes and time tagged by the RTUs.
In legacy SCADA systems, existing communication links between 
the control center and remote sites operate at very low speeds 
(could be on an order of 300bps to 9600bps). Note that present 
deployments of SCADA sysetms have variant models and technologies, 
which may have much better performances (for example, 61850-based systems).
Figure \ref{scadapic} describes a simple SCADA system.
In practice, more complicated SCADA system configurations exist.
Figure \ref{scadaconfig} lists three typical SCADA system
configurations (see, e.g., \cite{aga12}).

Recently, there have been several efforts to secure the national SCADA systems.
The examples are:
\begin{enumerate}
\item American Gas Association (AGA) \cite{aga12}. AGA is among the first to 
design cryptographic standard to protect SCADA systems.
American Gas Association (AGA) had originally been designing 
cryptographic standard to protect SCADA communication links and finished
the report AGA 12 part 1. The AGA 12 part 2 has been trransferred to 
IEEE P1711.
\item IEEE P1711 \cite{ieee1711}. This is transferred from AGA 12 part 2.
This standard effort tries to define a security protocol, 
the Serial SCADA Protection Protocol (SSPP), for control system serial 
communication. 
\item IEEE P1815. Standard for Electric Power Systems Communications -- 
Distributed Network Protocol (DNP3). The purpose of this standard 
is to document and make available the specifications for the DNP3 protocol.
\item IEC TC57 WG15 \cite{iectc57,iec608705}. IEC TC57 WG57 standardize
SCADA communication security via its IEC 608705 series.
\item NIST \cite{nist}. The NIST Industrial Control System Security (ICS)
group works on general security isseus related to control systems such as SACAD
systems.
\item National SCADA Test Bed Program \cite{testbed}.
The Department of Energy established the National Supervisory Control 
and Data Acquisition (SCADA) Test Bed program at Idaho National Laboratory 
and Sandia National Laboratory to ensure the secure, reliable and efficient 
distribution of power.
\end{enumerate}

\section{Threats to SCADA systems}
Several (real and simulated) attacks on SCADA systems were reported 
in the past few years \cite{ausattack,aurora,washingtonattack}.
In the Maroochy Shire attack \cite{ausattack},
an Australian man hacked into the Maroochy Shire, Queensland computerized 
waste management system and caused millions of liters of raw sewage to 
spill out into local parks, rivers and even the grounds 
of a Hyatt Regency hotel. It is reported that the 49-year-old Vitek 
Boden had conducted a series of electronic attacks on the Maroochy Shire 
sewage control system after his job application had been rejected. 
Later investigations found radio transmitters and computer equipments 
in Boden's car. The laptop hard drive contained software for accessing 
and controlling the sewage SCADA systems. The simulated 
Aurora attack \cite{aurora} conducted in March 2007 by the U.S. 
Department of Homeland Security resulted in the partial 
destruction of a \$1 million dollar large diesel-electric generator.

SCADA systems were not designed with public access in mind, they 
typically lack even rudimentary security. However, with the advent of 
technology and particularly the Internet, much of the technical information 
required to penetrate these systems is widely discussed in the public 
forums of the affected industries. Critical security flaws for SCADA 
systems are well known to potential attackers. It is feared that SCADA 
systems can be taken over by hackers, criminals, or terrorists. 
Some companies may assume that they use leased lines and therefore 
nobody has access to their communications. The fact is that it is 
easy to tap these lines \cite{tapline}. 
Similarly, frequency hopping spread spectrum radio and other wireless 
communication mechanisms frequently used to control remote terminal units
(RTU) can be compromised as well.  

Several efforts \cite{gao,nist,testbed} have been put on the analysis 
and protection of SCADA system security. 
According to these reports \cite{gao,nist,testbed}, 
the factors that have contributed
to the escalation of risk to SCADA systems include:
\begin{itemize}
\item The adoption of standardized technologies with
known vulnerabilities. In the past, proprietary hardware. software,
and network protocols made it difficult to understand how SCADA systems
operated---and therefore how to hack into them. Today, 
standardized technologies such as Windows, Unix-like operating systems,
and common Internet protocols are used by SCADA systems. Thus the 
number of people with knowledge to wage attacks on SCADA 
systems have increased.
\item The connectivity of control systems to other networks.
In order to provide decision makers with access to real-time information
and allowing engineers to monitor and control the SCADA systems from
different points on the enterprise networks, the SCADA systems are 
normally integrated into the enterprise networks. Enterprises are often
connected to partners' networks and to the Internet. Some enterprises
may also use wide area networks and Internet to transmit data to 
remote locations. This creates
further security vulnerabilities in SCADA systems.
\item Insecure remote connections.
Enterprises often use leased lines, wide area networks/Internet, and 
radio/microwave to transmit data between control centers and remote
locations. These communication links could be easily hacked.
\item The widespread availability of technical information about control
systems.
Public information about infrastructures and control systems is readily
available to potential hackers and intruders. For example,
Sean Gorman's dissertation (see, e.g., \cite{washingtonpost,rappaport})
mapped every business and industrial sector in the 
American economy to the fiber-optic network that connects them, using
materials that was available publicly on the Internet. In addition,
significant information on SCADA systems is publicly available 
(from maintenance documents, from former employees, and from 
support contractors, etc.). All these information could assist 
hackers in understanding the systems and to find ways to attack them.
\end{itemize}
Hackers may attack SCADA systems with one or more of the following actions.
\begin{enumerate}
\item Denial of service attacks by delaying or blocking the flow
of information through control networks. 
\item Make unauthorized changes to programmed instructions
in RTUs at remote sites, resulting in damage to equipment, premature shutdown
of processes, or even disabling control equipment.
\item Send false information to control system operators to disguise
unauthorized changes or to initiate inappropriate actions 
by system operators. 
\item Modify the control system software, producing
unpredictable results. 
\item Interfere with the operation of safety systems.
\end{enumerate}

The analysis in reports such as \cite{gao,nist,testbed} show 
that securing control systems 
poses significant challenges which include 
\begin{enumerate}
\item the limitations of current
security technologies in securing control systems. Existing Internet security 
technologies such as authorization, authentication, and encryption require
more bandwidth, processing power, and memory than control system components
typically have; Controller stations are generally designed to do 
specific tasks, and they often use low-cost, resource-constrained 
microprocessors;
\item the perception that securing control systems may not be economically
justifiable; and 
\item the conflicting priorities within organizations 
regarding the security of control systems.
In this paper, we will concentrate on the protection of SCADA remote 
communication links. In particular, we discuss the challenges on 
protection of these links and design new security technologies
to secure SCADA systems.
\end{enumerate}

\section{Securing SCADA remote connections}
Relatively cheap attacks could be mounted on SCADA system communication 
links between the control center and remote terminal units (RTU) since
there is neither authentication nor encryption on these links.
Under the umbrella of NIST ``Critical Infrastructure Protection 
Cybersecurity of Industrial Control Systems'', 
``American Gas Association (AGA) SCADA Encryption Committee'' 
has been trying to identify the functions and requirements for 
authenticating and encrypting SCADA communication links.
Their proposal \cite{aga12} is to build cryptographic
modules that could be invisibly embedded into existing SCADA systems 
(in particular, one could attach these cryptographic modules to modems 
of Figure \ref{scadaconfig}) so that all messages between modems are
encrypted and authenticated when necessary, and they have identified
the basic requirements for these cryptographic modules. However, 
due to the constraints of SCADA systems, no viable cryptographic protocols
have been identified to meet these requirements. In particular, the
challenges for building these devices are (see \cite{aga12}):
\begin{enumerate}
\item encryption of repetitive messages
\item minimizing delays due to cryptographic operations
\item assuring integrity with minimal latency
\begin{itemize}
\item intra-message integrity: if cryptographic modules buffer
message until the message authenticator is verified, it introduces
message delays that are not acceptable in most cases
\item inter-message integrity: reorder messages, replay messages, and 
destroy specific messages
\end{itemize}
\item accommodating various SCADA poll-response and retry strategies: delays
introduced by cryptographic modules may interfere with the SCADA system's
error-handling mechanisms (e.g., time-out errors)
\item supporting broadcast messages
\item incorporating key management
\item cost of device and management
\item mixed mode: some SCADA systems have cryptographic capabilities while
others not
\item accommodate to different SCADA protocols: SCADA devices are
manufactured by different vendors with different proprietary protocols.
\end{enumerate}

This paper designs efficient cryptographic mechanisms to address 
these challenges and to build cryptographic modules as 
recommended in \cite{aga12}.
These mechanisms can be used to build plug-in devices 
called sSCADA (secure SCADA) that could be inserted 
into SCADA networks so that 
all communication links are authenticated and encrypted. 
In particular, authenticated broadcast protocols are designed so that
they can be cheaply included into these devices. It has been a major
challenging task to design efficiently authenticated emergency broadcast 
protocols in SCADA systems.

The trust requirements in our security protocol design is as follows.
RTU devices are deployed in untrusted environments and 
individual remote devices could be controlled by adversaries.
The communication links are not secure but messages (maybe modified or 
re-ordered) could be delivered
to the destination with certain probability. In another word,
complete denial of service attacks (e.g., jamming) on the communication 
links are not addressed in our protocol. Compromising the control 
center in a SCADA system will make the entire system useless. Thus we 
assume that control centers are trusted in our protocol.

\section{sSCADA protocol suite}
\begin{center}
\begin{figure}[htb]
\centering{\includegraphics[scale=0.8]{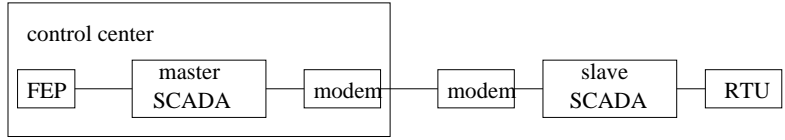}}
\caption{sSCADA with point-to-point SCADA configuration}
\label{secscadafig}
\end{figure}
\end{center}

The sSCADA protocol suite is proposed to overcome the challenges
that we have discussed in the previous section. 
sSCADA devices that are installed at the control center is called master 
sSCADA device, and sSCADA devices that are installed at remote sites are 
called slave sSCADA devices. Each master sSCADA device 
may communicate privately with several slave sSCADA devices. 
Once in a while, the master sSCADA device may also 
broadcast authenticated messages to several slave sSCADA devices 
(e.g., an emergency shutdown). An illustrative sSCADA device
deployment for point-to-point SCADA configuration is shown
in Figure \ref{secscadafig}.

\subsection{Vulnerabilities of a proposed protocol to AGA}
In this section, we discuss vulnerabilities of a proposed 
protocol to AGA. These analysis shows the challenges 
in designing secure communication protocols for SCADA systems.
A point to point secure channel protocol has been proposed 
by the AGA standard draft \cite{aga12,acns} (an open source
implementation could be found at \cite{scadasafe}). We first briefly
review this protocol in the following.

Preshared secrets are installed into the master sSCADA and 
slave sSCADA devices during deployment. These secrets are
used to negotiate session encryption and authentication keys
for the two devices. Each sSCADA device maintains a send sequence state 
variable in order to assign a sequence number to each ciphertext
message it sends. The send sequence variable is initialized to one at 
session negotiation, and incremented with every ciphertext
message sent. Let $i$ be the current send sequence number 
and $P=p_1\ldots p_n$ be the plaintext message that the sSCADA
device wants to send, where $p_j (j=1,\ldots, n)$ are blocks of 
the cipher block length (for example, if AES128 is used, then $p_j$ contains
128 bits). Then the sending sSCADA device enciphers
$P$ to the ciphertext $C$ as follows:
$$C=i c_1c_2\cdots c_n a$$
where
$$\begin{array}{l}
c_j=E_k[p_j\oplus E_k[i,j,00\ldots]],\\
a=\mbox{MAC}_{k'}[iP],
\end{array}$$
$E_k[\cdot]$ denotes the
encryption process using the key $k$, and $\mbox{MAC}_{k'}[\cdot]$
denotes the message authenticator computation process using the key $k'$.
The sending sSCADA device then sends $C$ to the receiving sSCADA device.
Let $\bar{C}=\bar{i}\bar{c_1}\bar{c_2}\cdots\bar{c_n}$ be the
message that the receiving sSCADA device receives.

At the receiving side, the sSCADA device maintains a receive sequence 
state variable in order to record the sequence number of the last 
authenticated message that it received. The receive sequence 
variable is initialized to zero at session negotiation. Before
decrypting the received ciphertext, the sSCADA device checks that the
sequence number $\bar{i}$ contained in the message is greater than
the sSCADA's receive sequence variable. If it is not, the sSCADA device 
discards the remainder of the message. This check is used to ensure
that an adversary cannot replay old messages (in the following,
our analysis shows that this protection could be easily defeated). 
Provided the sequence number check succeeds, the receiving sSCADA
device decrypts the message as follows:
$$\bar{P}=\bar{p_1}\bar{p_2}\cdots\bar{p_n}$$
where
$$\bar{p_j}=D_k[\bar{c_j}]\oplus E_k[\bar{i},j,00\ldots].$$
The receiving sSCADA device forwards the decrypted plaintext block
$\bar{p_j}$ to the SCADA system as soon as they are available. Finally,
the receiving sSCADA device computes the MAC for the message 
as follows:
$$\tilde{a}=\mbox{MAC}_{k'}[\bar{i}\bar{P}]$$
and compares it to the MAC $\bar{a}$. If the two match,
the sSCADA device updates its receive sequence variable to the sequence
number $\bar{i}$ of the received message, and otherwise it logs an error.

Now we present our attack on the above protocol in the following.
Assume that the adversary Carol controls the communication links between
the sending and receiving sSCADA devices and the current
receive sequence state variable at the receiving sSCADA side contains
the value $i_0$. When the sending sSCADA device sends the message
ciphertexts $C=i c_1c_2\cdots c_n a$ for $i>i_0$ in the future,
Carol forwards these ciphertexts to the receiving sSCADA device
by modifying one bit in the authenticators $a$ (all other
bits are forwarded as it is). When the receiving sSCADA devices
receives these ciphertexts, it checks the sequence numbers (which are
correct), decrypts the ciphertext blocks, and forwards the decrypted
plaintext blocks to the SCADA system. However, since the authenticators
have been tampered, the receiving sSCADA device fails to check
the authenticators. Thus the receiving sSCADA device will only log
errors without updating its receive sequence state variable. 
That is, the receiving sSCADA device will hold the value $i_0$ 
for its receive sequence state variable. At the same time, Carol
logs all these ciphertexts and observe what happens in the SCADA system.
Thus she can learn the meanings of these ciphertexts to the SCADA system.
At some time in the future, Carol wants the SCADA system to behave 
according to the ciphertext $C$ which contains a sequence number larger
than $i_0$. Carol can then just forward this ciphertext to the receiving
sSCADA device. Of course, Carol can also tamper the authenticator
so that the receiving sSCADA device still holds the value $i_0$ in its
receive state variable after processing this message. 
The receiving sSCADA device will just decrypt
this ciphertext and forward it to the SCADA system since the sequence
number contained in $C$ is larger than $i_0$. In another word, the SCADA
system is now in the complete control of Carol's hand.
 
Another potential pitfall 
in the proposed protocol is that same keys are used by two sides.
This leaves the door open for the attacker to replay the message
from one direction in the other direction. This vulnerability could
easily be fixed by using different padding schemes or using different
keys for different directions.
The original authors of the protocol has recommended some 
fix in \cite{scadasafe} to avoid our above attacks.

\subsection{Point-to-point secure channels}
In the previous section, we presented an attack on
the first draft of the AGA proposal. Though the protocol in the AGA proposal
could be fixed, in the following, we present a new secure solution.
In order to reduce the cost of sSCADA devices and management,
only symmetric key cryptographic techniques is used in our design.
Indeed, due to the slow operations of public key cryptography,
public key cryptographic protocols could introduce delays in message
transmission which are not acceptable to SCADA protocols.
Semantic security property \cite{goldwasser} is used to ensure that 
an eavesdropper has no information about the plaintext, even if it sees
multiple encryptions of the same plaintext. For example, even if the 
attacker has observed the ciphertexts of ``shut down'' and ``turn on'', 
it will not help the attacker to distinguish whether a new ciphertext
is the encryption of ``shut down'' or ``turn on''. In practice, the 
randomization technique is used to achieve this goal. For example, the message
sender may prepend a random string (e.g., 128 bits for AES-128) to 
the message and use special encryption modes such as chaining block cipher 
mode (CBC) or Hash-CBC mode (HCBC). In some mode, this random string is 
called the initialization vector (IV). This prevents information leakage from
the ciphertext even if the attacker knows several plaintext/ciphertext
pairs encrypted with the same key. 

Since SCADA communication links could be as low as 300bps and immediate
response are generally required, there is no sufficient 
bandwidth to send the random string (IV) each time with the ciphertext,
thus we need to design different cryptographic mechanisms to achieve
semantic security without additional transmission overhead. 
In our design, we use two counters shared between two communicating 
partners, one for each direction of communication. 

The counters are initially set to zeros and should be at least 128 bits, 
which ensures that the counter values will never repeat, avoiding replay 
attacks. The counter is used as the initialization vector (IV) in message 
encryptions if CBC or HCBC mode is used.  After each message encryption, 
the counter is increased by one if CBC mode is used and it is increased 
by the number of blocks of encrypted data if HCBC mode is used.
The two communicating partners are assumed to
know the values of the counters and the counters do not need to be
added to each ciphertext. Messages may get lost and the two 
counters need to be synchronized once a while (e.g., at off-peak time). 
A simple counter synchronization protocol is proposed for the 
sSCADA protocol suite. The counter synchronization protocol 
could also be initiated when some encryption/decryption errors 
appear due to unsynchronized counters.

In order for two sSCADA devices to establish a secure channel, 
a master secret key needs to be bootstrapped into the two devices
at the deployment time (or when a new sSCADA device is deployed
into the existing network).
For most configurations, secure channels are needed only between
a master sSCADA device and a slave sSCADA device. For some configurations,
secure channels among slave sSCADA devices may be needed also.
The secure channel identified with this master secret is used to 
establish other channels such as session secure channels, 
time synchronization channels, authenticated broadcast channels, 
and authenticated emergency channels.

Assume that ${\cal H}(\cdot)$ is a pseudorandom function (e.g., 
constructed from SHA-256)
and two sSCADA devices $A$ and $B$ share a secret 
${\cal K}_{AB}={\cal K}_{BA}$. Depending on the security policy, 
this key ${\cal K}_{AB}$ could be the shared master secret or a 
shared secret for one session which could be established from
the shared master key using a simple key establishment 
protocol (in order to achieve session key freshness, typically one 
node sends a random nonce to the other one and the other node sends 
the encrypted session key together with an authenticator on the 
ciphertext and the random nonce). Keys for different purposes 
could be derived from this secret as follows (it is not a good practice
to use the same key for different purposes). For example, 
$K_{AB}={\cal H}({\cal K}_{AB}, 1)$
is for message encryption from $A$ to $B$, 
$K'_{AB}={\cal H}({\cal K}_{AB}, 2)$
is for message authentication from $A$ to $B$, 
$K_{BA}={\cal H}({\cal K}_{AB}, 3)$
is for message encryption from $B$ to $A$, and 
$K'_{BA}={\cal H}({\cal K}_{AB}, 4)$
is for message authentication from $B$ to $A$.

Optional message authentication codes (MAC) are used for 
two parties to achieve 
data authentication and integrity. Message authentication codes 
that could be used for sSCADA implementation include 
HMAC \cite{hmac1,hmac2}, CBC-MAC \cite{desmode}, and others.
When party $A$ wants to send a message $m$ to party $B$ securely, $A$ computes
the ciphertext $c={\cal E}(C_A, K_{AB}, \bar{c_A}||m)$ and 
message authenticator $mac=MAC(K'_{AB}, C_A||c)$,
where $\bar{c_A}$ is the last $l$ bits of ${\cal H}(C_A)$
($l$ could be as large as possible
if bandwidth is allowed and 32 bits should be the minimal),
${\cal E}(C_A, K_{AB}, \bar{c_A}||m)$ denotes the encryption of $\bar{c_A}||m$
using key $K_{AB}$ and random-prefix (or IV) $C_A$ and $C_A$ is 
the counter value for the communication from $A$ to $B$. 
Then $A$ sends the following packets to $B$:
$$\begin{array}{ll}A\rightarrow B:& c,\  mac\ \mbox{(optional)}\end{array}$$
When $B$ receives the above packets, $B$ decrypts $c$, checks that 
$\bar{c_A}$ is correct, and verifies the message authenticator $mac$ 
if $mac$ is present. As soon as $B$ receives the first block
of the ciphertext, $B$ can check whether $\bar{c_A}$ is correct.
If it is correct, then $B$ continues the decryption and updates
it counter. Otherwise, $B$ discards the entire ciphertext.
If the message authenticator code $mac$ is present, $B$ also verifies
the correctness of $mac$. If $mac$ is correct, $B$ does nothing,
otherwise, $B$ may choose to inform $A$ that the message was corrupted
or try to re-synchronize the counters.

There are several implementation issues
on how to deliver the message to the target (e.g., RTU). For example,
we give a few cases in the following.
\begin{enumerate}
\item\label{case1} $B$ uses the counter to decrypt the first block 
of the ciphertext,
if the first $l$ bits of the decrypted plaintext is not consistent
with ${\cal H}(C_A)$, then the reason could be that the 
counter $C_A$ is not synchronized or that the ciphertext is corrupted.
$B$ may try several possible counters until the counter checking process 
succeeds. $B$ then uses the verified
counter and the corresponding key to decrypt the message and deliver
each block of the resulting message to the target as soon as it is 
available. If no counter could be verified 
in a limited number of trials. $B$ may notify $A$ of the transmission 
failure and initiate the counter synchronization protocol in the next 
section. The advantage of this implementation is that we have 
minimized delay from the cryptographic devices, thus minimize the
interference of SCADA protocols. Note that in this implementation,
the message authenticator $mac$ is not used at all.
If the ciphertext was tampered, we rely on the error correction 
mechanisms (normally CRC codes) in SCADA systems to discard 
the entire message. 
If CBC (respectively HCBC) mode is used, then the 
provable security properties (respectively, provable 
on-line cipher security properties) of CBC mode (respectively HCBC mode) 
\cite{bellarehcbc,bellarecbc}
guarantees that the attacker has no chance to tamper the ciphertext
so that the decrypted plaintext contains correct 
CRC that was used by SCADA protocols to achieve integrity. 
\item Proceed as in the above case \ref{case1}. In addition, 
the $mac$ is further checked
and the decrypted message is delivered to the SCADA system
only if the $mac$ verification passes. 
The disadvantage for this implementation is that these cryptographic
operations introduce significant delay for message delivery and it 
may interfere with SCADA protocols.
\item Proceed as in the above case \ref{case1}. 
The decrypted message is delivered
to the SCADA system as soon as they are available. After receiving the 
entire message and $mac$, $B$ will also verify $mac$. If the 
verification passes, $B$ do nothing. Otherwise, $B$ re-synchronizes
the counter with $A$ or initiates some other exception handling protocols.
\item In order to avoid delays introduced by cryptographic operations
and to check the $mac$ at the same time, 
sSCADA devices may deliver decrypted bytes immediately
to the target except the last byte. If the message authenticator $mac$ 
is verified successfully, the sSCADA device delivers the last byte 
to the target; Otherwise, the sSCADA device discards the last byte or 
sends a random byte to the target.
That is, we rely on the error correction mechanisms at the target
to discard the entire message. Similar mechanisms have been proposed in
\cite{aga12}. However, an attacker may insert garbages between
the ciphertext and $mac$ thus trick the sSCADA device to deliver
the decrypted messages to the SCADA system. If this happens, we essentially
do not get advantage from this implementation. Thus this implementation
is not recommended.
\item Instead of prepend $\bar{c_A}$ to the plaintext message,
one may choose to prepend three bytes of other specially formated
string to the plaintext message (three bytes bandwidth is normally 
available in SCADA systems) before encryption. This is an acceptable
solution though we still prefer our solution of prepending the hash outputs
of the counter.
\end{enumerate}
There could be other implementations to improve the performance and 
interoperability with SCADA protocols. sSCADA device should provide several
possible implementations for users to configure. Indeed, sSCADA devices 
may also be configured in a dynamic way that for different messages 
it uses different implementations.

In some SCADA communications, message authentication-only is sufficient.
That is, it is sufficient for $A$ to send $(m,\ mac)$ to $B$, 
where $m$ is the cleartext message and $mac=MAC(K'_{AB}, C_A||m)$. 
sSCADA device should provide configuration options to do message
authentication without encryption. In this case, even if the counter
value is not used as the IV, the counter value should still be authenticated
in the $mac$ and be increased after the operation. This will provide 
message freshness assurance and avoid replay attacks.
sSCADA should also support message pass-through mode. That is, message is 
delivered without encryption and authentication. In a summary, it should be
possible to configure an sSCADA device in such a way that some messages
are authenticated and encrypted, some messages are authenticated only,
and some messages are passed through directly. 

It is straightforward to show that our point-to-point secure channels
provide data authentication, data integrity, data confidentiality,
and weak data freshness (that is, messages arrive at the destination 
in the same order that was sent from the source).

\subsection{Counter synchronization}
In the point-to-point message authentication and encryption protocol, 
we assume that
both sSCADA devices $A$ and $B$ know each other's counter values $C_A$ and 
$C_B$.  In most cases, reliable communication in SCADA systems is 
provided and the security protocols in the previous section
work fine. Still we provide a counter synchronization protocol
so that sSCADA devices could synchronize their counters when necessary.
The counter synchronization protocol could be initiated by either 
side. Assume that $A$ initiates the counter synchronization protocol. Then the
protocol looks as follows:
$$\begin{array}{ll}
A\rightarrow B: & N_A\\
B\rightarrow A: & C_B, \ MAC(K'_{BA}, N_A||C_B)
\end{array}$$
This counter synchronization protocol is analogous to 
that in \cite{spins}.

The initial counter values of two sSCADA devices could be bootstrapped
directly. The above counter synchronization protocol could also be used
by two devices to bootstrap the initial counter values.
A master sSCADA device may also use the authenticated broadcast channel
that we will discuss in the next section to set several slave sSCADA 
devices' counters to the same value using one message. 

\subsection{Authenticated broadcast channels}
Encryption and authentication alone are not sufficient for SCADA applications.
For example, it is not acceptable to authenticate a message individually
in an emergent shutdown when timely responses from the RTU's are critical.
In order to support authenticated broadcast, we use one way key chains. 
This channel can be used to establish other 
channels such as authenticated emergency channels (see next section).

Typical authenticated broadcast channels require asymmetric cryptographic 
techniques,
otherwise any compromised receiver could forge messages from the sender.
Cheung \cite{cheung} proposed a symmetric cryptography based
source authentication technique in the context of authenticating 
communication among routers. Cheung's technique is based on delayed 
disclosure of keys by the sender. Later, it was used in the Guy Fawkes 
protocol \cite{anderson} for interactive unicast communication,
and in \cite{stream1,stream2,stream3,tesla1,tesla2} for streamed 
data multicast. Perrig, Szewczyk, Tygar, Wen, and Culler 
adapted delayed key disclosure based 
TESLA protocols \cite{tesla1,tesla2} to sensor networks for 
sensor broadcast authentication (the new adapted protocol is called 
$\mu$TESLA).
One-way key chains used in these protocols are analogous to the one-way
key chains introduced by Lamport \cite{lamport} and the S/KEY authentication
scheme \cite{skey}.

In the following, we briefly describe the authenticated 
broadcast scheme for SCADA systems.
At the sender (normally the master
sSCADA device or a computer connected to it) set up time,
the sender generates a one-way key
chain in the setup phase. In order to generate a one-way key chain of
length $n$, the sender chooses a random key ${\cal K}_n$ first,
then it applies the pseudorandom function ${\cal H}$ repeatedly to
${\cal K}_n$ to generate the remaining keys. In particular, for each
$i<n$, ${\cal K}_i={\cal H}({\cal K}_{i+1})$.

For the purpose of broadcast authentication, the sender 
splits the time into even intervals $I_i$. The duration of each time interval
is denoted as $\delta$ (e.g., $\delta=5$ seconds or $5$ minutes or even
2 hours), 
and the starting time of the interval $I_i$ is denoted as $t_i$. 
In another word, $t_i=t_0+i\delta$. 
At time $t_i$, the sender broadcasts the key ${\cal K}_i$.
Any device that has an authentic copy of key ${\cal K}_{i-1}$ can verify
the authenticity of the key ${\cal K}_i$ by checking whether
${\cal K}_{i-1}={\cal H}({\cal K}_i)$. Indeed, any device 
that has an authentic copy of some key ${\cal K}_{v}$ ($v<i$) can verify
the authenticity of key ${\cal K}_i$ since 
${\cal K}_v={\cal H}^{(i-v)}({\cal K}_i)$.

Let $d$ (a unit of time intervals) be the key disclosure delay factor. 
The value of $d$ is application dependent and could be configured at 
deployment time or after deployment (e.g., using 
the secure broadcast protocols itself). 
After $d$ is fixed, the sender
will use keying materials derived from key ${\cal K}_{i+d}$ to authenticate 
broadcast messages during the time interval $I_i$. Thus the message 
being broadcast during time interval $I_i$ could be verified by the 
receiver during the time interval $I_{i+d}$ after the sender broadcasts 
${\cal K}_{i+d}$ at time $t_{i+d}$. It is easy to see that
in order to achieve authenticity, the sender and the receiver need to be
loosely time synchronized. Otherwise, if the receiver time is slower than
the sender's time, an attacker can use published keys to impersonate the 
sender to the receiver. Typically the key disclosure delay should be 
greater than any reasonable round trip time between the sender and the
receiver. If the sender does not broadcast data frequently, 
the key disclosure delay may be significantly larger. For example, 
$d\delta$ could take the value of several hours for some SCADA systems.

If a receiver (typically a slave sSCADA device) is deployed at some time
during the interval $I_i$, the sender needs to bootstrap key  ${\cal K}_i$
on the one-way key chain to the receiver. The sender also needs
to bootstrap the key disclosure schedule which includes
the starting time $t_i$ of the time interval $I_i$, the key disclosure
delay factor $d$, and the duration $\delta$ of each time interval.
All these information could be bootstrapped to the receiver
using the point-to-point secure channel that we have 
designed in the previous section or using other channels such as manual
input. During a time interval $I_j\ (j>i)$, the receiver receives the 
broadcast key ${\cal K}_j$ from the sender and 
verifies  whether ${\cal K}_{j-1}={\cal H}({\cal K}_j)$.
If the verification is successful, the receiver updates its key
on the one-way key chain. If the receiver does not receive the broadcast
key during the time interval $I_j$ (either due to packet loss or due to
active denial of service attacks such as jamming attacks), 
it can update its key in the next time interval $I_{j+1}$.

When a receiver gets a packet from the sender, it first checks 
whether the key used for the packet authentication has been revealed.
If the answer is yes, then the attacker knows the key also and the 
packet could be a forged one. Thus the receiver needs to discard 
the packet. If the key have not been revealed yet, the receiver 
puts the packet in the buffer and checks the authenticity of the packet
when the corresponding key is revealed. As stated above, if the sender
and the receiver agree on the key disclosure schedule and the time is 
loosely synchronized, then message authenticity is guaranteed. However,
the protocol does not provide non-repudiation, that is, the receiver 
cannot convince a third party that the message was from the claimed sender.

If we assume that the time between  the sender and the receiver
are loosely synchronized and the pseudorandom function ${\cal H}(\cdot)$
and the message authentication code (MAC) are secure, then an analogous 
proof as in \cite{tesla2} could be used to show that the above 
authenticated broadcast channel is secure.
Note that we say that a {\em pseudorandom function} ${\cal H}(\cdot)$ 
is secure if  the function family $f_k(x)={\cal H}(k,x)$ 
is a pseudorandom function family in the sense of \cite{ggm} when $k$ is 
chosen randomly. That is, a function family $\{f_k(\cdot)\}$ is pseudorandom 
if the adversary with polynomially bounded resources cannot distinguish 
between a random chosen function from $\{f_k(\cdot)\}$ and a totally 
random function with non-negligible probability. 
We say that a message authentication scheme
MAC is secure if a polynomially bounded adversary will not succeed with 
non-negligible probability in the following game. A random $l$-bits key 
$k$ are chosen by the user. The adversary chooses messages 
$m_1,\ldots, m_t$ and the user generates the MAC codes on these messages 
using the key $k$.  The adversary succeeds if she could then generate 
a MAC code on a different message $m'\not= m_1,\ldots, m_t$.

Though the time synchronization between the sender and the receiver
plays an important role in the security of the protocol, they do not
need to have 100\% accurate clocks. If their clocks are sufficiently
accurate, then time synchronization protocol could be designed to synchronize
their clocks to meet the security requirements. The time synchronization
protocols could be based on the point-to-point secure channels 
discussed in the previous section.

\subsection{Authenticated emergency channels}
In our basic authenticated broadcast protocol, the receiver cannot 
verify the authenticity of the message immediately since it needs to
wait for the disclosure of the key after a time period of $d\delta$.
This is not acceptable for some broadcast messages such as an emergency 
shutdown. In order to overcome this challenge, the sender may
reveal the key used for emergency messages immediately or shortly after the
message broadcast. This will open the door for an adversary
to modify the emergency messages. For example, if the message passes through
a node $D$ before it reaches a node $C$, $D$ can discard the message
and create a different emergency message and forward it to $C$. 
In another case, an attacker may jam the target $C$ during the emergency 
broadcast period and sends $C$ a different emergency message (authenticated
using the revealed key for the emergency message) later.
However, these attacks are generally not practical since if the bad
guy jams the channel in a wireless environment, then he jams 
himself and he cannot receive
the authenticated broadcast message either.

\subsection{Authenticated emergency channels with finitely many messages}
In this section we design authenticated emergency channels
which can only broad finitely many emergency messages.
Assume that emergency messages are $e_1,\ldots, e_u$.
Without loss of generality, we may assume that 
$e_i=i$ for $i\le u$. Before the sender could authentically broadcast
these messages, it needs to carry out a commitment protocol.

Let $v$ be a fixed number. During the message commitment procedure, 
the sender chooses $v$ random numbers $N_1^i, \ldots, N_v^i$ for each 
$i\le u$. It then computes $r_{i,j}={\cal H}(e_i||N^i_j)$ for all 
$i\le u$ and $j\le v$.
Using the authenticated broadcast channel, the sender
broadcasts the commitments $\{r_{i,j}: i\le u \mbox{ and } j\le v\}$ 
to all receivers. Receivers store these commitments in their memory space.

Each time when the sender wants to broadcast the message $e_i$ to receivers
emergently,
it chooses a random unused $j\le v$, and broadcasts $(e_i, j, N_j^i)$ to 
all receivers. The receiver verifies that $r_{i,j}={\cal H}(e_i||N^i_j)$.
If the verification is successful, it knows that the message $e_i$ comes from
the sender and delivers it to the target. At the same time, the receiver
deletes the commitment $r_{i,j}$ from its memory space.

Note that after each message commitment procedure, the sender could
broadcast each message at most $v$ times. Thus the sender may decide to
initiate the message commitment protocol when any one of these messages has
been broadcast sufficiently many times (e.g., $v-1$ times). Each time
when the message commitment protocol is initiated, both the sender
and the receiver should delete all previous commitments from their
memory space.

The security of the emergency channel could be proved formally under the 
assumption that the pseudorandom function ${\cal H}(\cdot)$ is a secure
one-way function. That is, for any given $y$ with appropriate length,
one cannot find an $x$ such that ${\cal H}(x)=y$ with non-negligible 
probability.
\begin{theorem}
\label{emtheorem}
Assume that the authenticated broadcast channel is secure and
the pseudorandom function ${\cal H}(\cdot)$ is a secure one-way function.
Then the authenticity of messages that receivers accept from the 
emergency channel is guaranteed.
\end{theorem}

\noindent
{\em Sketch of Proof.} Assume for a
contradiction that the authenticity of the emergency protocol is broken.
That is, there is an adversary ${\cal A}$ who controls communication links and
manages to deliver a message $m$ to the receiver such that the sender
has not sent the message but the receiver accepts the message. We show
in the following that then ${\cal H}(\cdot)$ is not a secure one-way function.
Specifically, let $t$ be the total number of messages that the sender
can broadcast in the emergency channel with one commitment $\{r_{i,j}\}$, 
and $y_1,\ldots, y_t$ be $t$ randomly chosen strings with appropriate
lengths (i.e., they are potential outputs of ${\cal H}$).
We will construct an algorithm ${\cal P}$ that uses 
${\cal A}$ to compute a pre-image $x={\cal H}^{-1}(y_i)$ of some string
$y_i$ with non-negligible probability.

Since the broadcast channel is secure, we can always
assume that the commitment $\{r_{i,j}\}$ that the receivers accept 
are authentic. The algorithm ${\cal P}$ works by running ${\cal A}$ as follows.
Essentially, ${\cal P}$ simulates an authenticated broadcast channel for
${\cal A}$ with a sender $A$ and a receiver $B$.
\begin{enumerate}
\item ${\cal P}$ chooses a random number $l\le t$.
\item ${\cal P}$ computes a commitment $\{r_{i,j}\}$ as specified in
the emergency broadcast protocol.  ${\cal P}$ picks $t-l+1$ random values
from the  commitment $\{r_{i,j}\}$ and replace them with 
$y_l,y_{l+1},\ldots,y_t$.
\item ${\cal P}$ runs the sender's algorithm to 
authentically broadcast the modified commitment to $B$. 
\item For the first $l-1$ emergency messages,  ${\cal P}$  runs the 
sender's algorithm of the emergency broadcast protocol with no modification
to broadcast the pre-images of the $l-1$ unmodified commitments.
\item ${\cal P}$ then waits for ${\cal A}$ to deliver a fake message 
$x'$ that $B$ accepts as an authentic emergency broadcast.  
${\cal P}$ outputs $x'$ as one of the pre-images of $y_l,\ldots, y_t$.
\end{enumerate}
We briefly argue that ${\cal P}$ outputs the pre-image of 
one of the strings from $y_1,\ldots, y_t$ with non-negligible probability.
Since ${\cal A}$ succeeds with non-negligible probability in
convincing the receiver to accept a fake message, it must deliver
this message as the $l$-th message for some $l\le t$ in the authenticated 
emergency channel. Thus for this $l$, the algorithm ${\cal P}$ outputs 
a pre-image for one of 
the given strings with non-negligible probability.
\hfill{Q.E.D.}

\vskip 3pt
Theorem \ref{emtheorem} shows that messages received in the emergency
channel are authentic. However, it does not show whether these messages are
fresh. Indeed, when the sender broadcasts an emergency message
at the time $t$, the adversary may launch a denial of service attack
against the receiver or just does not deliver the message to the 
receiver. Thus the receiver will not be able to delete the commitment
of this message from its memory space. Later at time $t'$, the adversary 
delivers this message to the receiver and the receiver accepts it.
In our emergency channel, there is no way to avoid this kind of 
delayed message attacks. Thus when message freshness is important, 
one may use the revised authenticated emergency broadcast channel that 
we will discuss in the next section.

\subsection{Revised authenticated emergency channel}
There are basically two ways to guarantee the freshness of 
a received message. The first one is to use public key cryptography
together with time-stamps. The second solution is to let the receiver
send a nonce to the sender first and the sender authenticates
the message together with the nonce. As we have mentioned earlier,
public key cryptography is too expensive to be deployed in SCADA systems.
For the second solution, the delays introduced in nonce submission
process are generally not acceptable in an emergent situation.
In this section, we introduce a revised emergency broadcast protocol,
which provides weak freshness of received messages. Here weak freshness 
means that the received message is guaranteed to be in certain time limit 
$T$. In another word, at time $t$, the adversary cannot convince 
a receiver to accept a message that is posted before the time $t-T$.

Let the $u$ emergency messages be $e_1,\ldots, e_u$. Similar to 
the previous protocol, the sender needs to carry out a commitment protocol
before the authenticated emergency broadcast.
In the revised protocol, the sender chooses
$v$ random numbers $N_1^i, \ldots, N_v^i$ and $v$ expiration time points
$T_1^i<T_2^i<\ldots<T_v^i$ for each $i\le u$.
It then computes $r_{i,j}={\cal H}(e_i||N^i_j||T^i_j)$ for $i\le u$
and $j\le v$. Using the authenticated broadcast channel, the sender
broadcasts the commitments $\{r_{i,j}: i\le u \mbox{ and } j\le v\}$ 
to all receivers. Receivers store these commitments in their memory space.
The functionality of expiration time points in the revised protocol
is to guarantee that the commitment $r_{i,j}$ for the message $e_i$
expires at the time $T^i_j$. In another word, when the receiver
receives $(e_i, N^i_j,T^i_j)$, it will accepts the message only if
the current clock time of the receiver is earlier than $T^i_j$.

If the sender wants to send the message $e_i$ to receivers at time $t$,
it chooses a random unused $j\le v$ such that $t< T^i_j$, the estimated
transmission time from the sender to receiver is less than $T^i_j-t$,
and $T^i_j$ is the earliest time in the commitments that satisfies
these conditions. Then the sender broadcasts $(e_i, j, N_j^i, T_j^i)$ to 
all receivers. The receiver verifies that 
$r_{i,j}={\cal H}(e_i||N^i_j||T^i_j)$ and the current clock time 
of the receiver is earlier than $T^i_j$.
If the verification is successful, it knows that the message $e_i$ comes from
the sender and  delivers it to the target. At the same time it
deletes the commitment $r_{i,j}$ from its memory space. Otherwise,
the receiver discards the message.

The implementation of the revised emergency broadcast protocol
has the flexibility to choose the gaps between expiration time
points $T^i_j$s for each 
$i\le u$. The smaller the gap, the better the freshness property.
However, smaller gaps between $T^i_j$s add additional overhead
on the communication links. It is also possible, for different 
messages $e_i$, one chooses different values $v$. For example,
for more frequently broadcast message, the value of $v$ should be
larger. It is also important to guarantee that the commitment is
always sufficient and when only a few commitments are unused,
the sender should initiate a procedure for a new commitment.

The security of the revised emergency broadcast protocol can be proved
similarly as in Theorem \ref{emtheorem}. It is still possible for 
an adversary to delay an emergency message $(e_i, j, N_j^i, T_j^i)$
broadcast by the sender during the time period $[T_{j-1}^i, T_j^i]$ 
until $T_j^i$. However, she cannot delay the message to some 
time points after $T_j^i$. In another word, weak freshness of 
received messages are guaranteed in the revised authenticated 
emergency channel.

\section{Conclusion}
In this paper, we systemarically discussed the security issues
for SCADA systems and the challenges to design such a secure
SCADA system.  In particular, we present an attack on the protocols in 
the first version of AGA standard draft \cite{acns}. 
This attack shows that the security mechanisms
in the first draft of the AGA standard protocol could be easily defeated. 
We then proposed a suite
of security protocols optimized for SCADA/DCS systems which include:
point-to-point secure channels, authenticated broadcast channels, 
authenticated emergency channels, and revised authenticated 
emergency channels. These protocols are designed to address the 
specific challenges that SCADA systems have. 

Recently, there has been a wide interest for the secure design 
and implementation of smart grid systems \cite{doesmartgrid}.
SCADA system is one of the most important legacy systems of the
smart grid systems. Together with other efforts such as 
\cite{testbed,nist,ieee1711,ieee1815,iectc57,iec608705}, 
our work in this paper presents an inital step
for securing the SCADA section of the smart grid systems against cyber attacks.

\section*{Acknowledgement}
The author would like to thank the anonymouse referees for the 
excellent comments on the improvement of this paper.

\end{document}